\documentclass[aps,prl,reprint,amsmath,amssymb,nofootinbib]{revtex4-2}
\usepackage{graphicx}
\usepackage{bm}
\usepackage{xcolor}

\def\dashL{\mbox{\textbf{--~--~--}}}

\def\Lbox{\mbox{---}~{\hspace*{-.1in} $\square$}\hspace*{-.1in}~\mbox{---}}
\def\Lcirc{\mbox{---}~{\hspace*{-.1in}$\circ$}\hspace*{-.1in}~\mbox{---}}
\def\Ltriangle{\mbox{---}~{\hspace*{-.1in}$\triangle$}\hspace*{-.1in}~\mbox{---}}

\def\Lcross{\mbox{---}~{\hspace*{-.14in} $\times$}\hspace*{-.1in}~\mbox{---}}

\def\dashLcirc{\mbox{--~--}~{\hspace*{-.1in}$\circ$}\hspace*{-.1in}~\mbox{--~--}}
\def\dashLtriangle{\mbox{--~--}~{\hspace*{-.1in}$\triangle$}\hspace*{-.1in}~\mbox{--~--}}

\begin{document}

\title{Kolmogorov scale in turbulence of surface gravity waves}

\author{Zhou Zhang$^{*}$}
\author{Yulin Pan$^{\dagger}$}
\affiliation{
 Department of Naval Architecture and Marine Engineering, University of Michigan, Ann Arbor, Michigan 48109, USA
}

\begin{abstract}
In this paper, we study the analogue of the Kolmogorov scale in surface gravity wave turbulence, characterized by the cutoff wavenumber $k_c$ at which the power-law inertial range transitions into the dissipation range. We perform numerical simulations of the primitive dynamical equations with a broad-scale dissipation of magnitude $\gamma_0 k^2$ in spectral space to establish the relation between $k_c$ and $\gamma_0$. Our results show a scaling $k_c\sim\gamma_0^{\beta}$, where $\beta$ depends on the slope $\alpha$ of the power-law spectrum. We find that $\beta(\alpha)$ agrees more closely with the prediction obtained by balancing the nonlinear and dissipation terms in the dynamical equations than with that based on the kinetic equation. This observation reveals that non-resonant triad interactions play a more significant role than resonant quartet interactions in the formation of $k_c$.
\end{abstract}

\maketitle
\begingroup
\renewcommand{\thefootnote}{\fnsymbol{footnote}}
\footnotetext[1]{Corresponding author: joezhang@umich.edu}
\footnotetext[2]{Corresponding author: yulinpan@umich.edu}
\endgroup

\paragraph{Introduction.}---Turbulence refers to a state of dynamical systems involving interactions among many different scales. A general feature of turbulence is the formation of a power-law spectrum in the inertial range associated with a constant flux of, say, energy. For hydrodynamic turbulence, the power-law spectrum can be described by the $-5/3$ law, developed by Kolmogorov \cite{kolmogorov1941lst} based on a dimensional argument. For wave turbulence (i.e., a turbulent state formed by an ensemble of weakly nonlinear waves), the power-law spectrum can be found as an analytical solution of the wave kinetic equation \cite{zakharov2012kolmogorov,nazarenko2011wave}. The underlying theory, known as the wave turbulence theory (WTT), has found applications in many physical systems such as surface gravity waves \cite{hasselmann1962non,zakharov1965weak}, capillary waves \cite{zakharov1967weak}, internal gravity waves \cite{lvov2004energy,wu2023energy}, nonlinear optics \cite{dyachenko1992optical,picozzi2014optical}, and Majda-McLaughlin-Tabak models \cite{majda1997one,hrabski2024verification,simonis2024transition}.

As we go toward smaller scales, the inertial range of turbulence is replaced by the dissipation range and the spectrum departs from the power law. In hydrodynamic turbulence, the end of the inertial range is characterized by the Kolmogorov scale $\eta_K$. Based on Kolmogorov's hypothesis that the small-scale motions in turbulent flows are only dependent on the viscosity $\nu$ and the dissipation rate $\epsilon_K$, a dimensional analysis shows that $\eta_K$ is proportional to $\nu ^{3/4}$ \cite{pope2000turbulent} (or $\eta_K\sim Re^{-3/4}$ with $Re$ being the Reynolds number). Such a relation has been supported by experimental/numerical data \cite{ishihara2009study,vassilicos2015dissipation,cerbus2020small} and it implies that $\eta_K\rightarrow 0$ as $\nu\rightarrow 0$.

The counterpart of the Kolmogorov scale in wave turbulence is usually referred to as the cutoff wavenumber $k_c$ (with $k_c \sim 1/\eta_K$), which is much less studied. Existing theory \cite{kolmakov2004quasiadiabatic} postulates that $k_c$ can be determined by balancing the nonlinear interaction term and dissipation term in the wave kinetic equation (WKE). Based on this principle, a formula for $k_c$ in capillary wave turbulence has been derived, but experimental validation has yielded mixed results \cite{brazhnikov2001measurement,deike2012decay}. For surface gravity waves, this theory can be realized starting from the corresponding WKE:
\begin{equation}
    \frac{\partial n(\bm{k})}{\partial t}=I_n(\bm{k})+I_d(\bm{k}),
    \label{eq:kewithdissp}
\end{equation}
where $\bm{k}$ is the wave vector, $n$ is the wave action spectrum, and $I_n$ and $I_d$ are the nonlinear and dissipation terms respectively. The balance between $I_n$ and $I_d$ is equivalent to comparable magnitudes of the nonlinear timescale $\tau_n$ and the dissipation timescale $\tau_d$. The former can be estimated as the time for small perturbations to evolve against the background of a stationary spectrum \cite{falkovich1991nonstationary}, while the latter can be derived analytically from the dissipation formulation. For a power-law spectrum $n \sim k^{\mu}$ ($\mu=-4$ for the Kolmogorov-Zakharov solution) which corresponds to a 1D surface elevation spectrum $S_{\eta}\sim k^{\alpha}$ with $\alpha=\mu+3/2$, we can show (see Appendix A for details) that $I_n\sim k^{3\alpha+5}$ and $I_d\sim \gamma_0k^{\alpha+1/2}$, balancing which gives $k_c\sim \gamma_0^{\beta_k}$ with
\begin{equation}
    \beta_k(\alpha)= \frac{2}{4\alpha+9}.
    \label{eq:kcscalingkinetic}
\end{equation}
However, the mechanism regarding $k_c$ has never been studied for surface gravity waves, and the validity of the general theory based on the WKE is not clear. 

In this paper, we conduct the first numerical investigation of the mechanism underlying the formation of $k_c$ in surface gravity wave turbulence. Specifically, we simulate the primitive Euler equations incorporating a broad-scale dissipation term corresponding to $I_d$ in Eq.~\eqref{eq:kewithdissp}, until a stationary state is reached at which $k_c$ can be measured. The relation between $k_c$ and $\gamma_0$ is established directly from the numerical results, revealing an interesting finding summarized below.

We observe that with the increase of $\gamma_0$, a scaling $k_c\sim\gamma_0^{\beta}$ is approached. The exponent $\beta$ is found to be different from the prediction in Eq.~\eqref{eq:kcscalingkinetic} based on the kinetic theory. Instead, it lies much closer to $2/(\alpha-2)$, which can be obtained by balancing the nonlinear and dissipation terms in the dynamical Euler equations (which we will derive in detail). Such a balance in the dynamical equation is similar to the mechanism leading to the Kolmogorov scale $\eta_K\sim \nu ^{3/4}$ in hydrodynamic turbulence (considering that the Reynolds number computed from Kolmogorov length and velocity scales equals unity, i.e., $\eta_Ku_K/\nu=1$). Physically, the above observations imply that near $k_c$, non-resonant triad interactions play an important role in transferring energy to the dissipation range. While these non-resonant interactions are usually considered to be associated with reversible energy transfer, the situation can be different near $k_c$ where the transferred energy beyond $k_c$ is dissipated so that the transfer becomes irreversible. 

\paragraph{Numerical setup.}---We consider gravity waves on a 2D free surface of an incompressible, inviscid and irrotational fluid with infinite depth. The evolution of the wave field is governed by the Euler equations \cite{zakharov1968stability} for the surface elevation $\eta(x,y,t)$ and the velocity potential evaluated at the free surface $\psi(x,y,t)=\phi(x,y,z,t)|_{z=\eta}$: 
\begin{equation}
\begin{aligned}
    \frac{\partial\eta}{\partial t}=&-\nabla_{\bm{x}}\eta\cdot\nabla_{\bm{x}}\psi+(1+\nabla_{\bm{x}}\eta\cdot\nabla_{\bm{x}}\eta)\phi_z\\
    &+F^{-1}[-\gamma_0k^2\widehat{\eta}_{\bm{k}}],
\end{aligned}
\label{eq:eta}
\end{equation}
\begin{equation}
\begin{aligned}
    \frac{\partial\psi}{\partial t}=&-\eta-\frac{1}{2}\nabla_{\bm{x}}\psi\cdot\nabla_{\bm{x}}\psi+\frac{1}{2}(1+\nabla_{\bm{x}}\eta\cdot\nabla_{\bm{x}}\eta)\phi_z^2\\
    &+F^{-1}[-\gamma_0k^2\widehat{\psi}_{\bm{k}}],
\end{aligned}
\label{eq:psi}
\end{equation}
where $\nabla_{\bm{x}}=(\partial/\partial x, \partial/\partial y)$ denotes the horizontal gradient, and $\phi_z(x,y,t)=\partial\phi/\partial z|_{z=\eta}$ is the vertical velocity evaluated at the free surface. We model a broad-scale dissipation of magnitude $\gamma_0$ by the last terms in Eqs.~\eqref{eq:eta} and \eqref{eq:psi}, where $F^{-1}$ is the inverse Fourier transform, and $\widehat{\eta}_{\bm{k}}$ and $\widehat{\psi}_{\bm{k}}$ are the Fourier transforms of $\eta$ and $\psi$ respectively with $\bm{k}=(k_x,k_y)$ and $k=|\bm{k}|$. In Eq.~\eqref{eq:psi}, the density and gravitational acceleration both take the value of unity with a proper choice of mass and time units \cite{dommermuth1987high}.

The simulation of Eqs.~\eqref{eq:eta} and \eqref{eq:psi} is conducted by an order-consistent higher-order spectral (HOS) method \cite{west1987new} with the nonlinearity order $M=3$ to include up to cubic nonlinear terms, allowing for both triad and quartet interactions. For the time integration, we apply an integrating-factor scheme to solve the linear parts (without dissipation) analytically. The nonlinear and dissipation terms are integrated explicitly using a fourth-order Runge--Kutta method. Such a time-marching scheme has also been applied in our previous work \cite{zhang2025role} and a detailed description (in the context of capillary waves) can be found in \cite{pan2020high}.

We simulate a freely decaying wave field in a doubly periodic square domain with size $L=2\pi$ on each side, corresponding to a fundamental wavenumber $k_0=2\pi/L=1$. A spatial resolution of $1024\times 1024$ grid points is used, providing sufficient room for the variation of $k_c$ that is of interest to this study. The initial condition is set as an anisotropic ``truncated'' JONSWAP spectrum \citep{hasselmann1973measurements}, taking the form of $S(\omega,\theta)=D(\theta)J(\omega)H(\omega)$. Here, $\theta$ is the directional angle to the positive $x$ direction, and $D(\theta)$ is the directional spreading function:
\begin{equation}
    D(\theta)=
    \left\{
    \begin{array}{lc}
    \frac{2}{\pi}\cos^2\theta,& |\theta|\leq \pi/2\\
    0,& |\theta|>\pi/2
    \end{array},
    \right.
\label{eq:direction}
\end{equation}
and $J(\omega)$ is the JONSWAP spectrum with peak wavenumber $k_p=5$ and peak enhancement factor $\gamma=6$. $H(\omega)$ is a truncating function which takes the value of 1 for $\omega\leq k_a^{1/2}$ (with $k_a=45$) and zero otherwise. This truncation setup ensures that the spectral tail develops from quiescence, thus removing any possible dependence on the initial condition. The nonlinearity level of the wave field is characterized by the effective steepness $\epsilon=k_pH_s/2$ with $H_s$ being the significant wave height in the initial condition.

\begin{figure}
  \centerline{\includegraphics[scale =0.6]{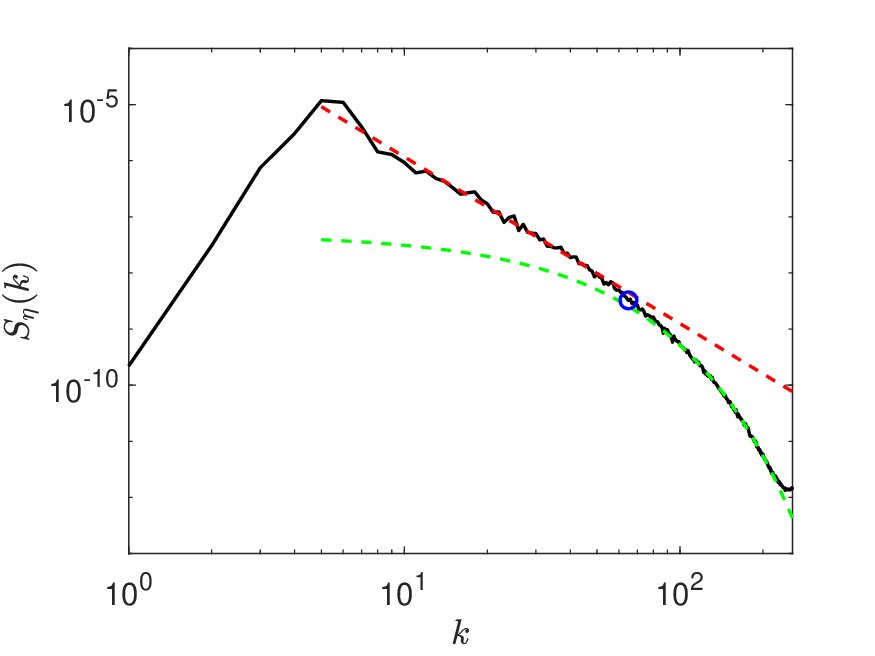}}
  \caption{A typical wave spectrum $S_{\eta}(k)$ ({\color{black}\rule[0.5ex]{0.5cm}{0.25pt}}) with the cutoff wavenumber $k_c$ ({\color{blue}$\circ$}) obtained from the power-law fit $S_{\eta}(k)\sim k^{\alpha}$ ({\color{red}\dashL}) and exponential fit $S_{\eta}(k)\sim \exp{(\xi k)}$ ({\color{green}\dashL}).}
\label{skkc}
\end{figure}

\paragraph{Results.}---We first present in Fig.~\ref{skkc} a typical result of the quasi-stationary spectrum obtained at $t=1800T_p$ (where $T_p$ is the peak period corresponding to $k_p$) with $\epsilon=0.0629$ and $\gamma_0=2\times 10^{-8}$. Here the spectrum is defined as the omnidirectional surface elevation spectrum $S_{\eta}(k)$ (or 1D energy spectrum), computed by
\begin{equation}
    S_{\eta}(k)=\int_0^{2\pi}\widehat{\eta}_{\bm{k}}\widehat{\eta}_{\bm{k}}^*kd\theta.
    \label{eq:sk}
\end{equation}
We see in Fig.~\ref{skkc} that an inertial range appears in $[k_p,k_c]$ with a power-law fit $S_{\eta}(k)\sim k^{\alpha}$, followed by a rapidly decaying dissipation range in $[k_c,k_{\max}]$ with an exponential fit $S_{\eta}(k)\sim \exp{(\xi k)}$. The cutoff wavenumber $k_c$ is then defined as the wavenumber at which the difference between the two fits is minimized (see a similar definition in Ref.~\cite{pan2015decaying}). 

\begin{figure}
  \centerline{\includegraphics[scale =0.6]{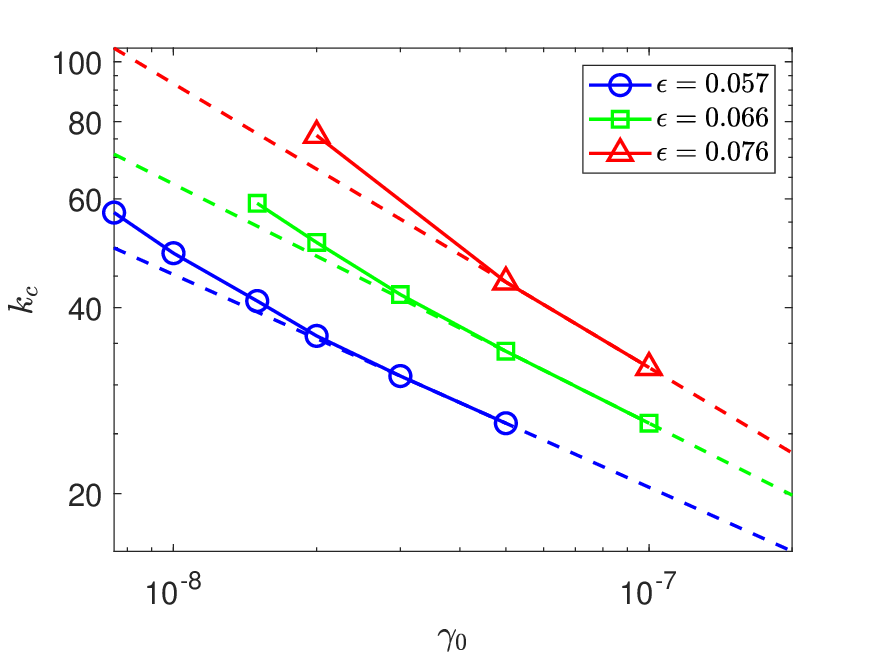}}
  \caption{The cutoff wavenumber $k_c$ as a function of $\gamma_0$ on a log-log scale with $\epsilon=0.057$ ({\color{blue}\Lcirc}), $\epsilon=0.066$ ({\color{green}\Lbox}), and $\epsilon=0.076$ ({\color{red}\Ltriangle}). The linear fits to the two data points with the largest $\gamma_0$ for each $\epsilon$ are marked by {\color{blue}\dashL}, {\color{green}\dashL}, and {\color{red}\dashL}, corresponding to slopes of $\beta=-0.34$, $\beta=-0.39$, and $\beta=-0.46$, respectively.}
\label{kcgamma}
\end{figure}

\begin{figure}
  \centerline{\includegraphics[scale =0.6]{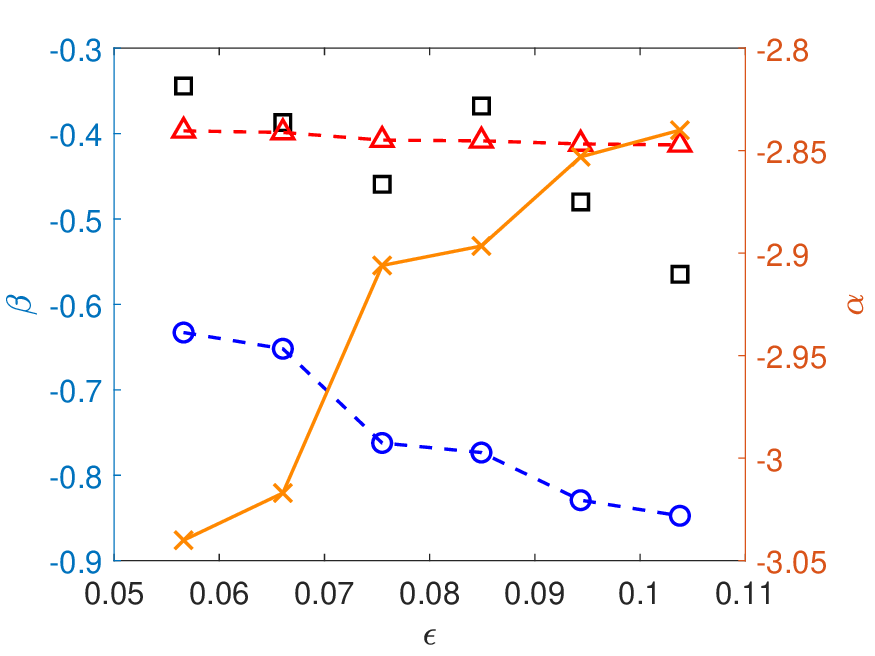}}
  \caption{The fitted exponent $\beta$ ({\color{black}$\square$}) with theoretical predictions $\beta_k(\alpha)$ ({\color{blue}\dashLcirc}) and $\beta_d(\alpha)$ ({\color{red}\dashLtriangle}), and the corresponding spectral slope $\alpha$ ({\color{orange}\Lcross}) as functions of $\epsilon$.}
\label{epsilonbeta}
\end{figure}

The relation between $k_c$ and $\gamma_0$ is plotted in Fig.~\ref{kcgamma} on a log-log scale at three nonlinearity levels $\epsilon=0.057$, $\epsilon=0.066$, and $\epsilon=0.076$. For all three nonlinearity levels, we see that $k_c$ increases with the decrease of $\gamma_0$, with a power-law scaling $k_c\sim\gamma_0^{\beta}$ achieved for relatively larger values of $\gamma_0$. The exponent $\beta$ at each nonlinearity level is computed by a linear fit of the two points with the largest $\gamma_0$. The slight upward deviation at smaller $\gamma_0$ is attributed to finite-$k_{\max}$ energy accumulation near the spectral boundary, and these points are excluded from the fit. Figure~\ref{epsilonbeta} compares the fitted values of $\beta$ with $\beta_k$ from Eq.~\eqref{eq:kcscalingkinetic}, i.e., the theoretical prediction based on the balance in the WKE. The corresponding spectral slope $\alpha$ as a function of $\epsilon$ is also shown in the figure. We see clear discrepancies between $\beta$ and $\beta_k$, indicating that new theory is needed to understand the cutoff wavenumber $k_c$.

We now describe the new theory based on the balance of nonlinear and dissipation terms in the dynamical equations, Eqs.~\eqref{eq:eta} and \eqref{eq:psi}, which aligns with the derivation for the Kolmogorov scale in hydrodynamic turbulence. Without loss of generality, let us start from Eq.~\eqref{eq:eta} (we have confirmed that using Eq.~\eqref{eq:psi} leads to the same result). We take the leading-order nonlinear term as $R_{n}(\bm{k})=\nabla_{\bm{x}}\eta\cdot\nabla_{\bm{x}}\psi$, and the dissipation term as $R_{d}(\bm{k})=F^{-1}[-\gamma_0k^2\widehat{\eta}_{\bm{k}}]$. An order-of-magnitude analysis in spectral space shows
\begin{equation}
\begin{aligned}
    R_n(\bm{k})=\nabla_{\bm{x}}\psi\cdot\nabla_{\bm{x}}\eta &\sim O(H_k/T_k)\cdot O(H_k/L_k)\\ 
    &\sim O(H_k^2T_k^{-1}L_k^{-1}),
\end{aligned}
\label{eq:n2scaling0}
\end{equation}
\begin{equation}
    R_d(\bm{k})=F^{-1}[-\gamma_0k^2\widehat{\eta}_{\bm{k}}]\sim O(\gamma_0 k^2 H_k),
    \label{eq:dscaling}
\end{equation}
where $H_k$, $T_k$ and $L_k$ are the characteristic wave height, period and length at scale $k$, respectively. In particular, we have
\begin{equation}
    O(L_k)\sim O(k^{-1}),
    \label{eq:Lk}
\end{equation}
\begin{equation}
    O(T_k)\sim O(k^{-1/2}),
    \label{eq:Tk}
\end{equation}
\begin{equation}
    O(H_k)\sim O(k^{(\alpha-1)/2}),
    \label{eq:Hk}
\end{equation}
where the first two equations follow from the dispersion relation, and the last one results from considering the spectrum $S_{\eta}(k)\sim k^{\alpha} \sim O(H_k^2k)$. At scale $k_c$, we expect that $ R_n(\bm{k})/R_d(\bm{k})\sim O(1)$. Substitution of Eqs.~\eqref{eq:Lk}, \eqref{eq:Tk}, and \eqref{eq:Hk} into Eqs.~\eqref{eq:n2scaling0} and \eqref{eq:dscaling} gives $k_c\sim \gamma_0^{\beta_d}$ with 
\begin{equation}
    \beta_d(\alpha)=\frac{2}{\alpha-2}.
    \label{eq:kcscalingdynamic}
\end{equation}

A plot of $\beta_d$ from Eq.~\eqref{eq:kcscalingdynamic} is added in Fig.~\ref{epsilonbeta}. We see that the numerical result of $\beta$ lies between $\beta_k$ and $\beta_d$, but is much closer to the latter. This observation is consistent with the breakdown of the kinetic description at high wavenumbers. For gravity waves, the characteristic nonlinear timescale $\tau_n\sim k^{-3/2}$ \cite{falkovich1991nonstationary,newell2011wave}, implying increasingly strong nonlinear interactions at large $k$, where the timescale-separation condition $\tau_n/T_k\gg1$ underlying the kinetic equation eventually ceases to hold. The primitive dynamical equations thus provide a more appropriate description of the spectral tail. Within this framework, the observed scaling further indicates that non-resonant triad interactions play a more significant role than resonant quartet interactions in shaping the spectrum close to $k_c$. We note that non-resonant triad interactions are usually associated with reversible energy transfer (i.e., bound modes \cite{zhang2022numerical}), except when they are connected to form resonant quartets \cite{zhang2025role}. Here the mechanism does not arise from the resonant quartets formed by connected triads, because otherwise $\beta=\beta_k$ would be expected. Instead, dissipation renders the energy transfer by non-resonant triads irreversible at the spectral tail. Specifically, bound modes generated above $k_c$ are instantly dissipated, so that energy is drained through non-resonant interactions coupled with dissipation. This mechanism is analogous to the formation of the Kolmogorov scale in hydrodynamic turbulence, but differs from the previous understanding based on the balance in the WKE.

\paragraph{Conclusions.}---In this paper, we present an investigation into the mechanism underlying $k_c$, i.e., the counterpart of the Kolmogorov scale, in deep-water surface gravity wave turbulence. For relatively large dissipation magnitude $\gamma_0$, our numerical results show that the scaling exponent $\beta$ lies between theoretical predictions based on the balance of nonlinear and dissipation terms in the kinetic and dynamical equations, but is much closer to the latter. This fact indicates that non-resonant triad interactions (instead of the previously assumed quartet resonances) play a dominant role in the formation of $k_c$, analogous to the Kolmogorov scale in hydrodynamic turbulence. 
\bibliography{references}

\section*{End Matter}
\appendix
\setcounter{equation}{0}
\renewcommand{\theequation}{A\arabic{equation}}
\paragraph{Appendix A: Derivation of $\beta_k(\alpha)$ from the WKE.}---Our objective is to obtain the scaling $k_c\sim \gamma_0^{\beta_k}$ based on the condition $I_n(\bm{k})/I_d(\bm{k})\sim O(1)$. We first consider the nonlinear term in the WKE which is expressed in the form of a collision integral \cite{zakharov2012kolmogorov,nazarenko2011wave}:
\begin{equation}
\begin{aligned}
    I_n(\bm{k})=4\pi\iiint_{-\infty}^{\infty}& |T_{0123}|^2 n(\bm{k})n(\bm{k}_1)n(\bm{k}_2)n(\bm{k}_3) \\
    &\left[\frac{1}{n(\bm{k})}+\frac{1}{n(\bm{k}_1)}-\frac{1}{n(\bm{k}_2)}-\frac{1}{n(\bm{k}_3)}\right] \\
    &\delta(\bm{k}+\bm{k}_1-\bm{k}_2-\bm{k}_3) \\
    &\delta(\omega_k+\omega_1-\omega_2-\omega_3) \\
    &d\bm{k}_1d\bm{k}_2d\bm{k}_3,
\end{aligned}
\label{eq:inkinetic}
\end{equation}
where $T_{0123}=T(\bm{k},\bm{k}_1,\bm{k}_2,\bm{k}_3)$ is the interaction coefficient, $\omega_k\sim |\bm{k}|^{1/2}$ is the dispersion relation and $\delta$ is the Dirac delta function.

We exploit the homogeneity property of the interaction coefficient $T_{0123}$ \cite{zakharov2012kolmogorov}:
\begin{equation}
    T(\bm{k},\bm{k}_1,\bm{k}_2,\bm{k}_3)=k^3T(\bm{k}/k,\bm{k}_1/k,\bm{k}_2/k,\bm{k}_3/k),
    \label{eq:intcoeff}
\end{equation}
and the scaling property of the Dirac delta function:
\begin{equation}
    \delta(\bm{k}+\bm{k}_1-\bm{k}_2-\bm{k}_3)=k^{-2}\delta(\bm{k}/k+\bm{k}_1/k-\bm{k}_2/k-\bm{k}_3/k),
    \label{eq:deltak}
\end{equation}
\begin{equation}
    \delta(\omega_k+\omega_1-\omega_2-\omega_3)=k^{-1/2}\delta(1+\omega_1/\omega_k-\omega_2/\omega_k-\omega_3/\omega_k).
    \label{eq:deltaomega}
\end{equation}
Based on the assumed power-law spectrum
\begin{equation}
    n(\bm{k})\sim k^{\alpha-3/2},
\label{eq:nkscaling}
\end{equation}
we have
\begin{equation}
\begin{aligned}
    n(\bm{k})n(\bm{k}_1)n(\bm{k}_2)n(\bm{k}_3) &\left[\frac{1}{n(\bm{k})}+\frac{1}{n(\bm{k}_1)}-\frac{1}{n(\bm{k}_2)}-\frac{1}{n(\bm{k}_3)}\right]\\
    &\sim k^{3(\alpha-3/2)}.
\end{aligned}
\label{eq:nproduct}
\end{equation}
Then we consider the dimension of the integral in Eq.~\eqref{eq:inkinetic}. For an arbitrary function $f(\bm{k},\bm{k}_1,\bm{k}_2,\bm{k}_3)$, we have
\begin{equation}
\begin{aligned}
    &\iiint_{-\infty}^{\infty} f(\bm{k},\bm{k}_1,\bm{k}_2,\bm{k}_3)d\bm{k}_1d\bm{k}_2d\bm{k}_3 \\
    =&k^6\iiint_{-\infty}^{\infty} f(\bm{k},\bm{k}_1,\bm{k}_2,\bm{k}_3)d(\bm{k}_1/k)d(\bm{k}_2/k)d(\bm{k}_3/k).
\end{aligned}
\label{eq:integral}
\end{equation}
Substituting Eqs.~\eqref{eq:intcoeff}, \eqref{eq:deltak}, \eqref{eq:deltaomega}, and \eqref{eq:nproduct} into Eq.~\eqref{eq:inkinetic} and making use of the property in Eq.~\eqref{eq:integral}, we can obtain the dimensional relation
\begin{equation}
    I_n(\bm{k})\sim k^6k^{-2}k^{-1/2}k^{3(\alpha-3/2)}k^6\sim k^{3\alpha+5}.
    \label{eq:icscaling}
\end{equation}

The dissipation term arises from the terms $-\gamma_0k^2\widehat{\eta}_{\bm{k}}$ and $-\gamma_0k^2\widehat{\psi}_{\bm{k}}$ in the evolution equations for $\widehat{\eta}_{\bm{k}}$ and $\widehat{\psi}_{\bm{k}}$ respectively. We consider the energy density $e_{\bm{k}}$ defined as 
\begin{equation}
    e_{\bm{k}}=\frac{1}{2}(k\widehat{\psi}_{\bm{k}}\widehat{\psi}_{\bm{k}}^* + \widehat{\eta}_{\bm{k}}\widehat{\eta}_{\bm{k}}^*).
\label{eq:ek}
\end{equation}
The contribution from dissipation terms to the time derivative of $e_{\bm{k}}$ is calculated by
\begin{equation}
    \begin{aligned}
        \frac{\partial e_{\bm{k}}}{\partial t} &= \frac{1}{2}\{ k[(-\gamma_0k^2\widehat{\psi}_{\bm{k}})\widehat{\psi}_{\bm{k}}^*+(-\gamma_0k^2\widehat{\psi}_{\bm{k}}^*)\widehat{\psi}_{\bm{k}}] + \\
        &\qquad\quad [(-\gamma_0k^2\widehat{\eta}_{\bm{k}})\widehat{\eta}_{\bm{k}}^*+(-\gamma_0k^2\widehat{\eta}_{\bm{k}}^*)\widehat{\eta}_{\bm{k}}]\} \\
        &= (-\gamma_0k^2)(k\widehat{\psi}_{\bm{k}}\widehat{\psi}_{\bm{k}}^* + \widehat{\eta}_{\bm{k}}\widehat{\eta}_{\bm{k}}^*) \\
        &= -2\gamma_0k^2 e_{\bm{k}}.
    \end{aligned}
\end{equation}
Then we can obtain the dissipation term $I_d(\bm{k})$ in the WKE:
\begin{equation}
    I_d(\bm{k}) = \frac{\partial (e_{\bm{k}}/\omega_k)}{\partial t} = -2\gamma_0k^2e_{\bm{k}}/\omega_k = -2\gamma_0k^2 n(\bm{k}).
    \label{eq:dndtdissipation}
\end{equation}
Substituting Eq.~\eqref{eq:nkscaling} into Eq.~\eqref{eq:dndtdissipation}, we have
\begin{equation}
    I_d(\bm{k})\sim \gamma_0 k^{\alpha+1/2}.
    \label{eq:idscaling}
\end{equation}
Finally, the kinetic scaling $k_c\sim \gamma_0^{\beta_k}$ can be derived by combining Eqs.~\eqref{eq:icscaling} and \eqref{eq:idscaling} with the condition $I_n/I_d\sim O(1)$, which gives
\begin{equation}
    \beta_k(\alpha)=\frac{2}{4\alpha+9}.
\end{equation}

\end{document}